\documentclass[debug]{rmaa}

\usepackage{paralist}

\usepackage{psfrag,color}

\usepackage[latin1]{inputenc}

\title{Comparative analysis of sky quality and meteorological variables during the total lunar eclipse on 14-15 April 2014 and their effect on qualitative measurements of the Bortle scale} 

\author{
  C. G\'oez Ther\'an\altaffilmark{1,2,3} 
  and S. Vargas Dom\'inguez\altaffilmark{4}}

\altaffiltext{1}{CINDES Research Group, Department of Engineering, Universidad Libre.}
\altaffiltext{2}{Research Group CENIT, Universidad Nacional de Colombia.}
\altaffiltext{3}{Olympiades Office, Astronomy and Astrophysics, Universidad Antonio Nari\~no.}
\altaffiltext{4}{Universidad Nacional de Colombia - Sede Bogot\'a - Facultad de Ciencias - Observatorio Astron\'omico Nacional.}

\shortauthor{G\'oez Ther\'an \& Vargas Dom\'inguez}
\shorttitle{Sky quality during a total lunar eclipse}

\fulladdresses{
\item Cristian G\'oez Ther\'an: CINDES Research Group, Department of Engineering, Universidad Libre, Bogotá,
Colombia (cristian.goezt@unilibre.edu.co). 
\item Santiago Vargas Dom\'inguez: Universidad Nacional de Colombia - Sede Bogot\'a - Facultad de Ciencias - Observatorio Astron\'omico Nacional - Carrera 45 \# 26-85, Bogot\'a - Colombia (svargasd@unal.edu.co).}

\listofauthors{G\'oez Ther\'an \& Vargas Dom\'inguez}
\indexauthor{G\'oez Ther\'an, C.}
\indexauthor{Vargas Dom\'inguez, S.}

\abstract{A total lunar eclipse is plausible to have an influence on the variation of some environmental physical parameters, specifically on the conditions of the sky brightness, humidity and temperature. During the eclipse on 14$^{th}$-15$^{th}$  April 2014, these parameters were measured through a photometer and a weather station. The  obtained results allow the comparison, practically, of the optimal conditions for observational astronomy work in the Tatacoa desert and therefore to certify it as suitable perfect place to develop night sky astronomical observations. This investigation determined, to some extent, the suitability of this place to carry out astronomical work and research within the optical range. Thus, the changes recorded during the astronomical phenomenon allowed the classification of the sky based on the Bortle Scale.}

\resumen{Es factible que un eclipse total de Luna tenga influencia en la variaci\'on de par\'ametros f\'isicos en una zona ambiental, espec\'ificamente en el brillo del cielo, humedad y temperatura. Durante el eclipse del 14 y 15 de abril de 2014, estos par\'ametros se midieron a trav\'es de un fot\'ometro y de una estaci\'on meteorol\'ogica. Los resultados obtenidos permiten hacer comparaciones de manera pr\'actica sobre las condiciones \'optimas para el trabajo de astronom\'ia observacional en el desierto de la Tatacoa (Colombia) y as\'i catalogarlo como un lugar apto para realizar observaciones astron\'omicas nocturnas. Esta investigaci\'on permite determinar hasta cierto punto, la idoneidad de este lugar para llevar a cabo trabajos astron\'omicos e investigaciones dentro del rango \'optico. De esta manera, los cambios registrados durante el fen\'omeno astron\'omico permitieron la clasificaci\'on del cielo con base en la escala de Bortle.}

\addkeyword{atmospheric effects}
\addkeyword{eclipses}
\addkeyword{methods: observational}
\addkeyword{site testing}

\begin{document}
\maketitle

\section{Introduction and Motivation}
\label{sec:intro}

Through the history of our species, the universe populated with stars, galaxies and other celestial objects has been visible in the darkness of the night sky, inspiring questions about the cosmos and our relationship to it, and therefore star gazing has been crucial not only for astronomy but also for literature, arts, philosophy and multiple human activities. Nevertheless, the technological advances of our society added to world population growth resulted in light pollution in most living areas where people do not have the opportunity to enjoy the night sky. Furthermore it has also been demonstrated issues on human health caused by artificial lighting and impacts on the behavior of plants and animals. For years now, astronomers have highlighted the negative consequences of the current situation and are promoting the conservation of the dark sky through multiple initiatives \citep{DarkSkies}. Millions of astronomy enthusiasts will benefit from these efforts that were not an issue for our ancestors.

In Colombia, the interest in astronomy dates back to the end of the 17th century according to some recent evidence found in a historical manuscript describing some ideas about the universe from Antonio S\'anchez de Cozar, a humble priest whose work, drafted  between 1676 and 1696, can be considered as the first original study of astronomy written in Colombia  \citep{Portilla&Moreno}. Some time later, Jos\'e Celestino Mutis, a big pioneer of science and knowledge, leaded a botanic expedition and founded an astronomical observatory as part of his initiative to spread science in the region. Erected in 1803, the National Astronomical Observatory of Colombia is the first astronomical observatory that was built in the Americas. The work by \citet{AriasdeGreiff} details the different historical and cultural aspects related to the development of astronomy in Colombia until the 20th century.

Even though the sky conditions for astronomical observations in the optical range are not very suitable in Colombia, there is a great interest from amateur astronomers, and still quite a few places in which people can enjoy the wonders of the night sky. Many careers of professional astronomers are the result of their {\bf youth} interest in astronomy while being part of amateur groups. This was the case of the geographer, meteorologist and astrometry specialist, William Cepeda Pe\~na,  an important disseminator of astronomy since 1965.  In the 70s he began astronomical studies in Colombia, such as a star occultation by the Moon (1988-1995), the movement of the Sun using the Hipparcos catalog (2002) and, in particular, related to eclipses observations, e.g.,  annular solar eclipse (1995), annular eclipse (1980), total solar eclipse (1998), works grouped in \citet{Cepeda}.

With a significant valuable legacy behind, we continue investigating on eclipses observed from Colombia. In this work, we report on information collected, analyzed and interpreted during the total lunar eclipse on April 14$^{th}$ and 15$^{th}$ in 2014, one of the four total lunar eclipses visible during 2014 and 2015 from Colombia. Particularly, we complement the work by \citet{Goez&Vargas} doing a deeper analysis of the data acquired from the Tatacoa Desert, a valuable natural location in the country, by measuring meteorological variables and the sly brightness to study the variation of these parameters during the occurrence of the eclipse. This work presents a statistical study and comparative analysis of sky quality to establish qualitative measurements of the Bortle scale.

\section{Methodology}
\label{sec:errors}

\begin{figure}[!t]
  \includegraphics[width=\columnwidth]{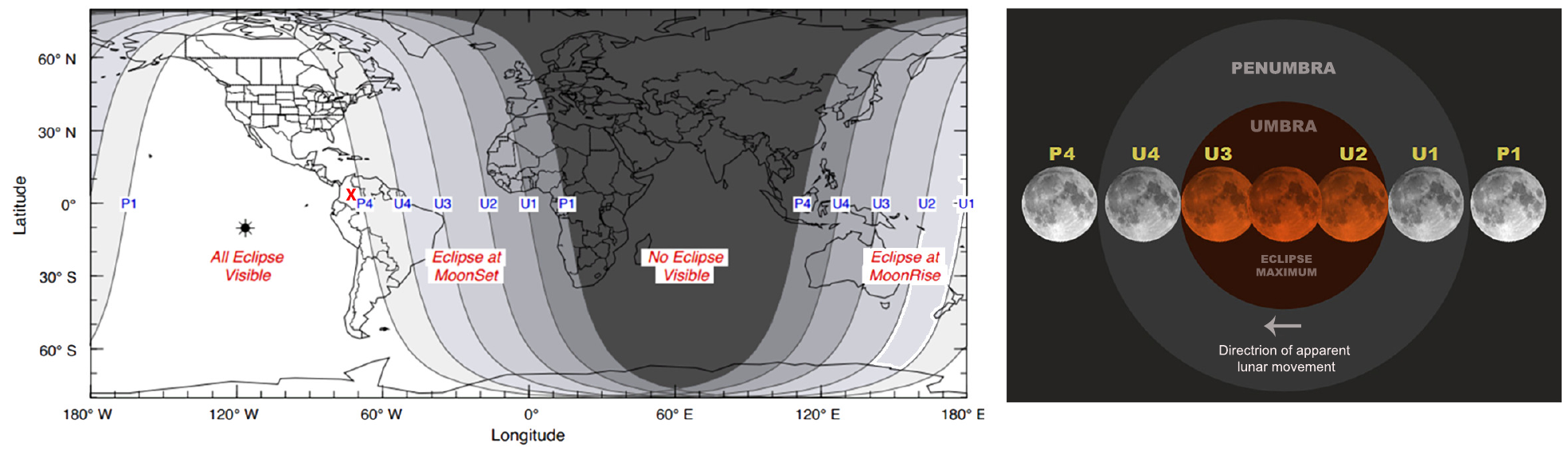}
  \caption{Visibility map of the total lunar eclipse on 14$^{th}$-15$^{th}$ April 2014 (left panel). The red X marks the location in Colombia where the eclipse was observed to carry out this research work. All different stages of a total lunar eclipse are also shown in the figure (right panel). Modified from NASA Reference Publication 1178.}
  \label{fig:1}
\end{figure}

An eclipse takes place when three celestial bodies, i.e. the Earth, the Sun and the Moon, line up or get closer to be. A total lunar eclipse is a well-known astronomic event that occurs when planet Earth stands in the way between the Sun and the Moon, allowing our natural satellite to enter into the cone of shadow that casts the Earth, thus, getting darker and turning into a characteristic russet color during the total occultation. At the totality phase, the Earth obstructs most of the solar rays that arrive at the Moon, which has to happen during full-moon phase, but some portion (mainly the red part of the visible light spectrum) is deflected by the Earth's atmosphere and hits the lunar surface making the moon getting a brownish-red color from the typical yellow light. The effect and dye depends also on the atmospheric conditions of our planet (clouds, dust particles, clouds of gas due to the volcanic eruptions, fires and other gas emissions close to the location where the eclipse observation is carried out), and on the distance between the Moon and the center of the umbra.  Figure~\ref{fig:1} shows the visibility map for the total lunar eclipse on 14$^{th}$-15$^{th}$ April 2014  and the different phases of this type of eclipse, which is far more common than a total solar eclipse. The total lunar eclipse was followed from the astronomical observatory of Tatacoa and other places in Colombia, with monitoring that depended terribly on the local meteorological conditions from the observing points.

In this work we aim at measuring the variations of sky quality \citep{Cinzano} and main meteorological variables during the total lunar eclipse on April 14$^{th}$ and 15$^{th}$ of 2014 and their effect on qualitative measurements of the Bortle scale. Tracking changes in environmental parameters and sky brightness prior to and during the eclipse occurrence are used to classify the sky according to the official global scales.

\section{Observation and measurements}
\label{sec:observations}

\subsection{Location}
\label{sec:place}

\begin{figure}[!t]
  \includegraphics[width=\columnwidth]{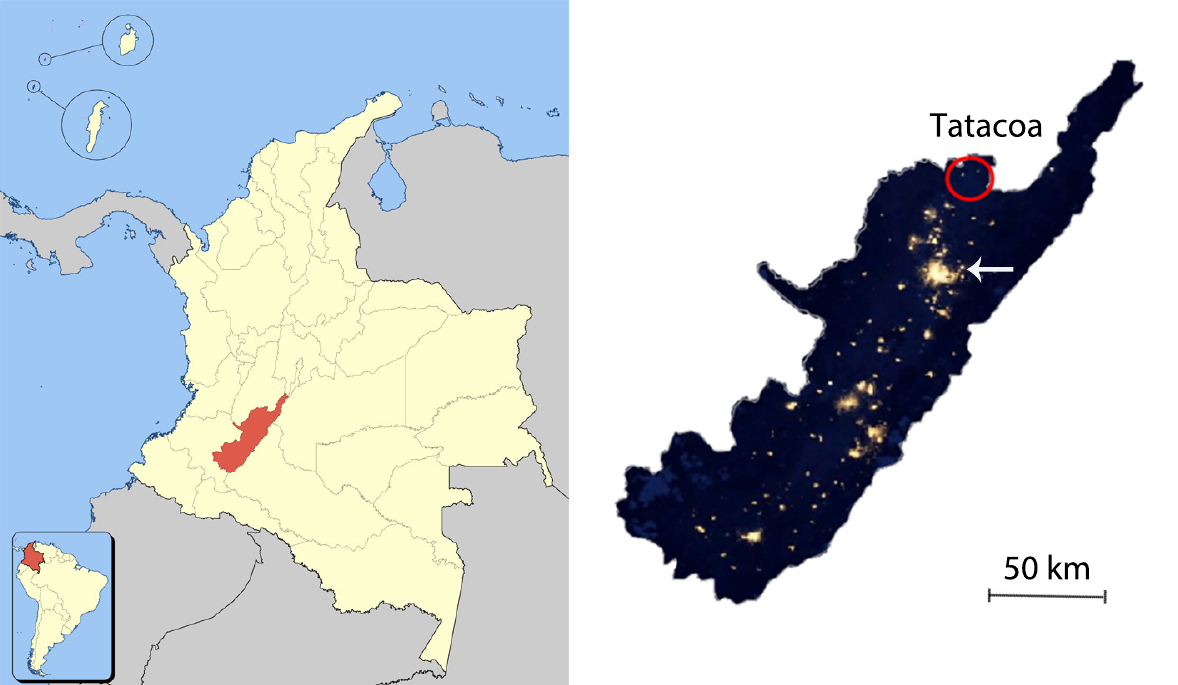}
  \caption{Map of Colombia with a colored reddish area highlighting the Department of Huila (left panel) and a night sky satellite image (right panel) showing some illuminated areas in the same Department corresponding to small towns and the city of Neiva (white arrow). The location of Tatacoa Desert is encircled in red.}
  \label{fig:2}
\end{figure}

The observation of the total lunar eclipse on 14$^{th}$-15$^{th}$ April, 2014 was carried out from the astronomical observatory of the Tatacoa, near the town of Villavieja in the Department of Huila, as shown in figure~\ref{fig:2}. The geographic coordinates of this location are 3$^{\circ}$14$^\prime$ North and 75$^{\circ}$10$^\prime$ West.

The Tatacoa Desert is one of the most exotics landscape in Colombian geography with an area of 370 km$^2$. This dry tropical forest is the second largest arid zone in the country after the Guajira Peninsula, with geomorphic principally of {\it estoraques} and {\it cavarcas}, among others. We decided to select this place for the follow up of the eclipse, motivated by the success of previous visits pursuing astronomical observations of the Milky Way, globular clusters, nebulae and meteor showers, besides the reasonably good average conditions of the location in terms of clear nights, low clouds and water vapor \citep{Goez,Goez&Vargas,Pinzon2016}.

\subsection{Calibration}
\label{sec:calibration}

{\bf Before} the acquisition of scientific data, the equipment was tested and prepared. The installation, assembling and calibration process of the different devices and sensors began at 00:40 UTC, i.e.,  photometers, weather stations, telescopes, computers and CCD cameras, in order to ensure an optimal and reliable data collection, before the eclipse begin. Figure~\ref{fig:3} displays the main components of the equipment.

\subsection{Data}
\label{sec:measurements}

Data acquisition for scientific measurements started right after the first contact, P1, at 4:53:37 UTC. Contact times and all different phases of the eclipse are listed in Table~\ref{tab:1}. Measurements of sky brightness were taken pointing the SQM to the zenith in all cases.

\begin{table}[!t]\centering
  \setlength{\tabnotewidth}{0.5\columnwidth}
  \tablecols{3}
  \setlength{\tabcolsep}{2.8\tabcolsep}
  \caption{Stages of the total lunar eclipse} \label{tab:1}
 \begin{tabular}{cl}
    \toprule
    STAGE& \multicolumn{1}{c}{Time (UTC)} \\
    \midrule
    P1 & 04:53:37   \\
    U1 & 05:58:19\\
    U2 & 07:06:47  (Beginning of Totality Phase)\\
   MAX & 07:45:40 (Totality)\\
    U3 & 08:24:35 (Completion of Totality Phase\\
    U4 & 09:33:04\\
    P4 & 10:37:37 \\
    \bottomrule
  \end{tabular}
\end{table}

\begin{figure}[!t]
  \includegraphics[width=\columnwidth]{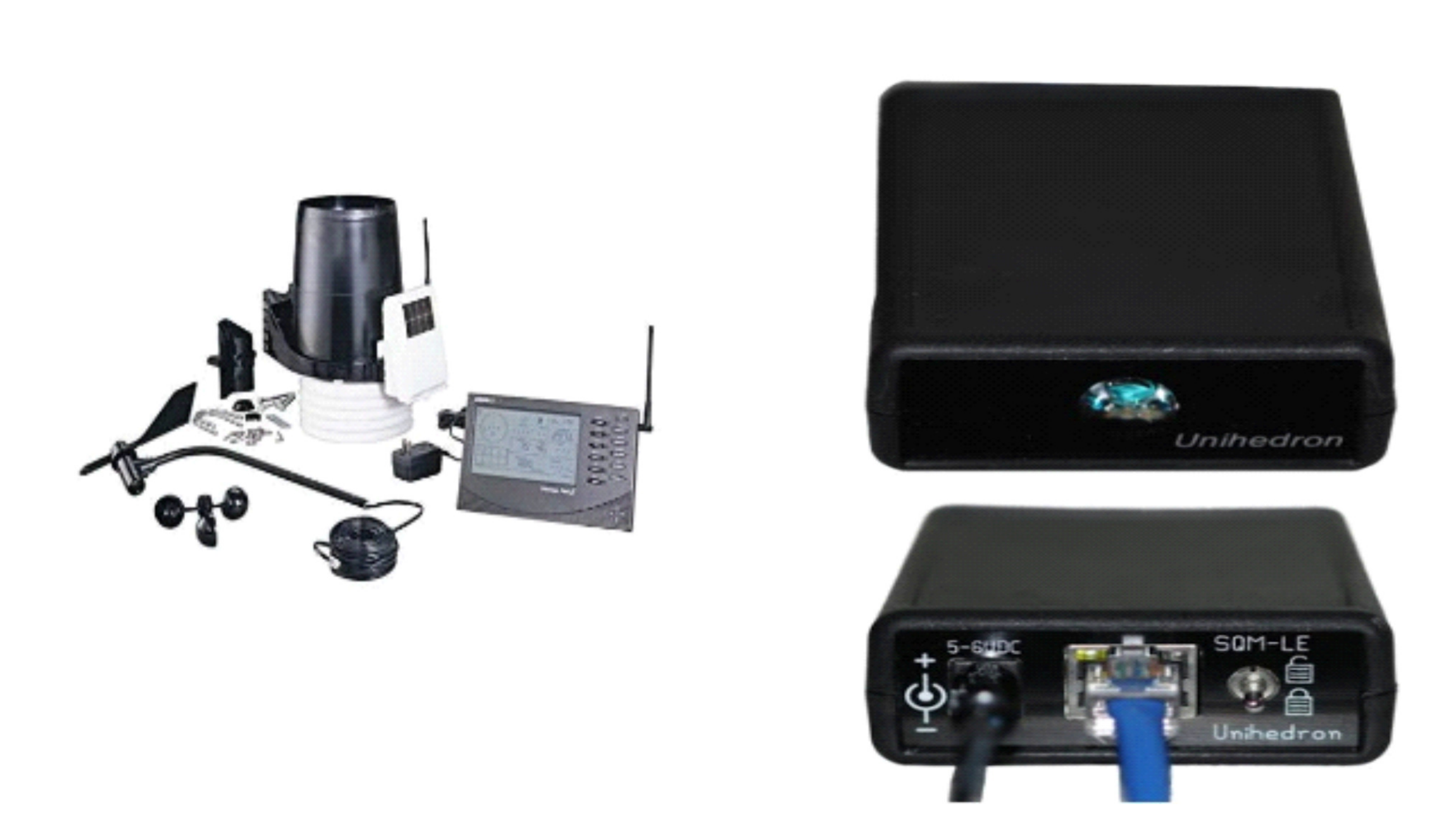}
  \caption{Main equipment and sensors that were used to measure sky quality and meteorological variables. See the text for details.}
  \label{fig:3}
\end{figure}

We use a photometer called SQM-LE to measure the brightness of the sky\footnote{www.unihedron.com/projects/darksky/}, a weather station, WMR200 Davis Instruments Pro with wireless sensors, as shown in figure~\ref{fig:3}. Furthermore, three CCD cameras for Celestron astrophotography were employed in order to record all stages of the eclipse  (P1, U1, U2, U3 and U4) by using a calibrated Meade ETX-90 telescope (two-star method) and with a permanent monitoring of the Moon to obtain the alt-azimuth values during the data collection period.

\begin{figure}[!t]
  \includegraphics[width=\columnwidth]{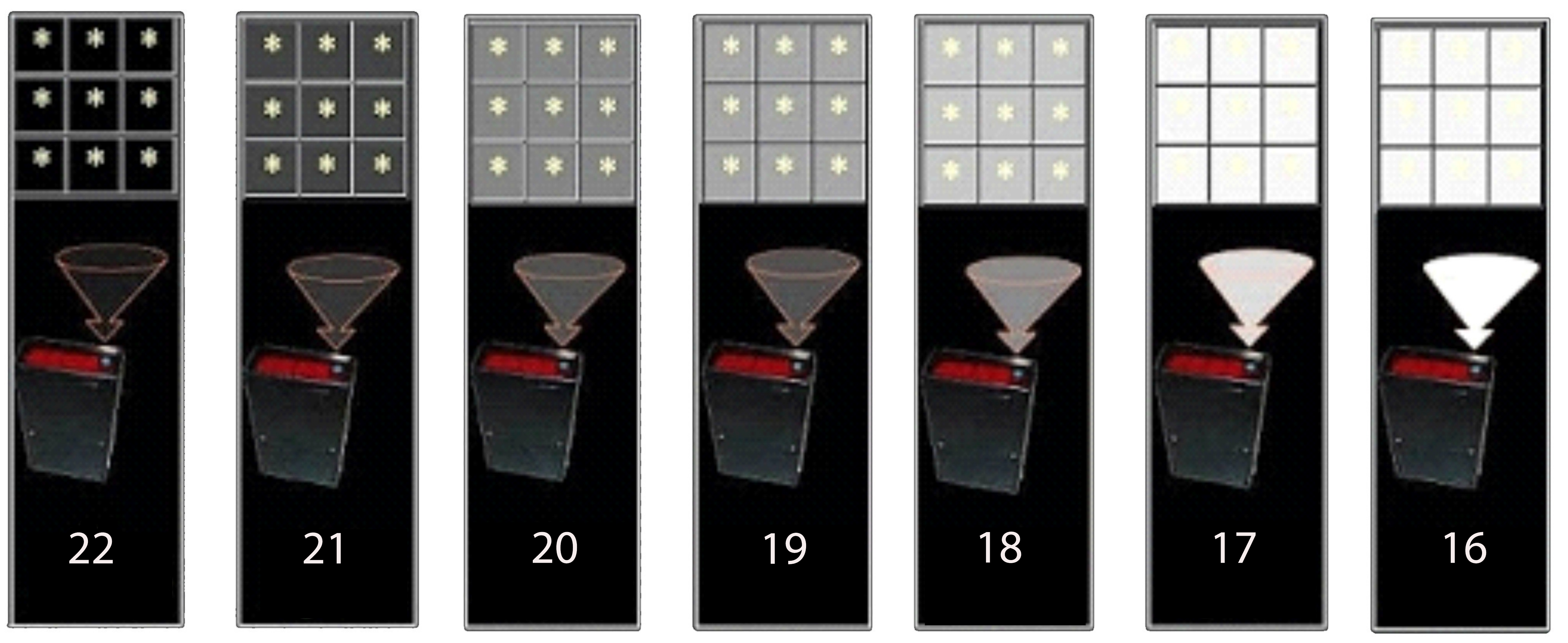}
  \caption{Sketch showing the use of a sky quality meter (SQM) to measure the amount of light striking the sensor. Numbers in every panel correspond to the value of sky brightness measured in units of mag. arcsec$^{-2}$ characterizing the quality of the sky. Taken from www.nightwise.org}  
  \label{fig:4}
\end{figure}

The photometer acquires data on a scale of mag.arsec$^{-2}$. Figure~\ref{fig:4} exemplifies the use of this instrument and the way we refer to the magnitude as describing the brightness of an object, i.e. the amount of light striking the sensor. Sky brightness, humidity, pressure and horizontal coordinates of the Moon were registered every 5 minutes and tabulated for the subsequent statistical analysis using the software package SPSS \citep{Argyrous}.

Table~\ref{tab:2} presents all the collected data during the observing session.

\begin{table}[]
\centering
\caption{Data} \label{tab:2}
\resizebox{\textwidth}{!}{%
\begin{tabular}{ccccccccc} 
\hline
DATA & LOCAL TIME & UTC  & SQM (mag.arcsec$^{-2}$)   & TEMPERATURE ($^\circ$C) & PRESSURE (mbar) &MOON HEIGHT ($^\circ$) & HUMIDITY ($\%$)  & STAGE                          \\
\hline
1     & 19:40      & 0:40 & 17.11 & 32.3                & 1000    & 26.6              & 43      & CALIBRATION OF EQUIPMENT AND SENSORS  \\
2     & 19:45      & 0:45 & 17.04 & 32.2                & 1000    & 27.8              & 43      & CALIBRATION OF EQUIPMENT AND SENSORS  \\
3     & 19:50      & 0:50 & 16.94 & 32                  & 1000    & 28.9              & 43      & CALIBRATION OF EQUIPMENT AND SENSORS  \\
4     & 19:55      & 0:55 & 16.92 & 31.9                & 1000    & 30.1              & 43      & CALIBRATION OF EQUIPMENT AND SENSORS  \\
5     & 20:00      & 1:00 & 16.95 & 31.8                & 1000    & 31.3              & 43      & CALIBRATION OF EQUIPMENT AND SENSORS  \\
6     & 20:05      & 1:05 & 16.9  & 31.8                & 1000    & 32.3              & 43      & CALIBRATION OF EQUIPMENT AND SENSORS  \\
7     & 20:10      & 1:10 & 16.82 & 31.7                & 1001    & 33.7              & 43      & CALIBRATION OF EQUIPMENT AND SENSORS  \\
8     & 20:15      & 1:15 & 17    & 31.6                & 1001    & 34.9              & 43      & CALIBRATION OF EQUIPMENT AND SENSORS  \\
9     & 20:20      & 1:20 & 17.2  & 31.6                & 1001    & 36                & 43      & CALIBRATION OF EQUIPMENT AND SENSORS  \\
10    & 20:25      & 1:25 & 17.15 & 31.6                & 1001    & 37.2              & 43      & CALIBRATION OF EQUIPMENT AND SENSORS  \\
11    & 20:30      & 1:30 & 17.18 & 31.5                & 1001    & 38.4              & 43      & CALIBRATION OF EQUIPMENT AND SENSORS  \\
12    & 20:35      & 1:35 & 17.03 & 31.5                & 1001    & 39.6              & 43      & CALIBRATION OF EQUIPMENT AND SENSORS  \\
13    & 20:40      & 1:40 & 16.72 & 31.3                & 1001    & 40.8              & 43      & CALIBRATION OF EQUIPMENT AND SENSORS  \\
14    & 20:45      & 1:45 & 16.5  & 31.2                & 1001    & 41.9              & 43      & CALIBRATION OF EQUIPMENT AND SENSORS  \\
15    & 20:50      & 1:50 & 16.35 & 31.1                & 1001    & 43.1              & 43      & CALIBRATION OF EQUIPMENT AND SENSORS  \\
16    & 20:55      & 1:55 & 16.29 & 31                  & 1002    & 44.3              & 43      & CALIBRATION OF EQUIPMENT AND SENSORS  \\
17    & 21:00      & 2:00 & 16.71 & 31                  & 1002    & 45.4              & 43      & CALIBRATION OF EQUIPMENT AND SENSORS  \\
18    & 21:05      & 2:05 & 16.76 & 30.9                & 1002    & 46.6              & 42      & CALIBRATION OF EQUIPMENT AND SENSORS  \\
19    & 21:10      & 2:10 & 16.73 & 30.8                & 1002    & 47.8              & 42      & CALIBRATION OF EQUIPMENT AND SENSORS  \\
20    & 21:15      & 2:15 & 16.6  & 30.6                & 1002    & 48.9              & 42      & CALIBRATION OF EQUIPMENT AND SENSORS  \\
21    & 21:20      & 2:20 & 16.51 & 30.5                & 1002    & 50.1              & 42      & CALIBRATION OF EQUIPMENT AND SENSORS  \\
22    & 21:25      & 2:25 & 16.49 & 30.2                & 1003    & 51.2              & 42      & CALIBRATION OF EQUIPMENT AND SENSORS  \\
23    & 21:30      & 2:30 & 16.39 & 30                  & 1003    & 52.4              & 42      & CALIBRATION OF EQUIPMENT AND SENSORS  \\
24    & 21:35      & 2:35 & 16.39 & 29.8                & 1003    & 53.5              & 41      & CALIBRATION OF EQUIPMENT AND SENSORS  \\
25    & 21:40      & 2:40 & 16.27 & 29.7                & 1003    & 54.7              & 41      & CALIBRATION OF EQUIPMENT AND SENSORS  \\
26    & 21:45      & 2:45 & 16.27 & 29.6                & 1003    & 55.8              & 41      & CALIBRATION OF EQUIPMENT AND SENSORS  \\
27    & 21:50      & 2:50 & 15.52 & 29.6                & 1003    & 56.9              & 41      & CALIBRATION OF EQUIPMENT AND SENSORS  \\
28    & 21:55      & 2:55 & 15.55 & 29.6                & 1003    & 58                & 41      & CALIBRATION OF EQUIPMENT AND SENSORS  \\
29    & 22:00      & 3:00 & 15.57 & 29.6                & 1003    & 59.2              & 41      & CALIBRATION OF EQUIPMENT AND SENSORS  \\
30    & 22:05      & 3:05 & 15.52 & 29.7                & 1003    & 60.3              & 41      & CALIBRATION OF EQUIPMENT AND SENSORS  \\
31    & 22:10      & 3:10 & 15.79 & 29.7                & 1003    & 61.4              & 41      & CALIBRATION OF EQUIPMENT AND SENSORS  \\
32    & 22:15      & 3:15 & 15.91 & 29.8                & 1003    & 62.4              & 41      & CALIBRATION OF EQUIPMENT AND SENSORS  \\
33    & 22:20      & 3:20 & 15.84 & 29.8                & 1003    & 63.5              & 41      & CALIBRATION OF EQUIPMENT AND SENSORS  \\
34    & 22:25      & 3:25 & 15.57 & 29.9                & 1003    & 64.6              & 41      & CALIBRATION OF EQUIPMENT AND SENSORS  \\
35    & 22:30      & 3:30 & 15.50 & 29.9                & 1003    & 65.6              & 41      & CALIBRATION OF EQUIPMENT AND SENSORS  \\
36    & 22:35      & 3:35 & 15.44 & 30.0                & 1003    & 66.6              & 41      & CALIBRATION OF EQUIPMENT AND SENSORS  \\
37    & 22:40      & 3:40 & 15.38 & 30.0                & 1003    & 67.6              & 42      & CALIBRATION OF EQUIPMENT AND SENSORS  \\
38    & 22:45      & 3:45 & 15.32 & 30.0                & 1003    & 68.6              & 42      & CALIBRATION OF EQUIPMENT AND SENSORS  \\
39    & 22:50      & 3:50 & 15.26 & 30.1                & 1003    & 69.6              & 42      & CALIBRATION OF EQUIPMENT AND SENSORS  \\
40    & 22:55      & 3:55 & 15.20 & 30.1                & 1003    & 70.5              & 42      & CALIBRATION OF EQUIPMENT AND SENSORS  \\
41    & 23:00      & 4:00 & 15.14 & 30.2                & 1003    & 71.4              & 41      & CALIBRATION OF EQUIPMENT AND SENSORS  \\
42    & 23:05      & 4:05 & 15.08 & 30.2                & 1003    & 72.2              & 42      & CALIBRATION OF EQUIPMENT AND SENSORS  \\
43    & 23:10      & 4:10 & 15.02 & 30.3                & 1003    & 73.1              & 42      & CALIBRATION OF EQUIPMENT AND SENSORS  \\
44    & 23:15      & 4:15 & 14.95 & 30.3                & 1003    & 73.8              & 41      & CALIBRATION OF EQUIPMENT AND SENSORS  \\
45    & 23:20      & 4:20 & 14.89 & 30.4                & 1003    & 74.5              & 41      & CALIBRATION OF EQUIPMENT AND SENSORS  \\
46    & 23:25      & 4:25 & 14.83 & 30.4                & 1003    & 75.1              & 42      & CALIBRATION OF EQUIPMENT AND SENSORS  \\
47    & 23:30      & 4:30 & 14.77 & 30.5                & 1003    & 75.7              & 42      & CALIBRATION OF EQUIPMENT AND SENSORS  \\
48    & 23:35      & 4:35 & 14.71 & 30.5                & 1003    & 76.1              & 42      & CALIBRATION OF EQUIPMENT AND SENSORS  \\
49    & 23:40      & 4:40 & 14.65 & 30.5                & 1003    & 76.5              & 43      & CALIBRATION OF EQUIPMENT AND SENSORS  \\
50    & 23:45      & 4:45 & 14.59 & 30.6                & 1003    & 76.8              & 43      & CALIBRATION OF EQUIPMENT AND SENSORS  \\
51    & 23:50      & 4:50 & 14.53 & 30.6                & 1003    & 76.9              & 43      & P1                                    \\
52    & 23:55      & 4:55 & 14.47 & 30.7                & 1003    & 76.9              & 43      & P1                                    \\
53    & 0:00       & 5:00 & 14.40 & 30.7                & 1003    & 76.9              & 43      & P1                                    \\
54    & 0:05       & 5:05 & 14.34 & 30.8                & 1003    & 76.7              & 43      & P1                                    \\
55    & 0:10       & 5:10 & 14.28 & 30.8                & 1003    & 76.4              & 43      & P1                                    \\
56    & 0:15       & 5:15 & 14.22 & 30.9                & 1003    & 76                & 43      & P1                                    \\
57    & 0:20       & 5:20 & 14.16 & 30.9                & 1003    & 75.5              & 43      & P1                                    \\
58    & 0:25       & 5:25 & 14.10 & 31.0                & 1003    & 74.9              & 43      & P1                                    \\
59    & 0:30       & 5:30 & 14.04 & 31.0                & 1003    & 74.3              & 43      & P1                                    \\
60    & 0:35       & 5:35 & 13.98 & 31.0                & 1003    & 73.6              & 43      & P1                                    \\
61    & 0:40       & 5:40 & 13.91 & 31.1                & 1003    & 72.8              & 43      & P1                                    \\
62    & 0:45       & 5:45 & 13.85 & 31.1                & 1003    & 72                & 43      & P1                                    \\
63    & 0:50       & 5:50 & 14.28 & 28.1                & 1003    & 71.1              & 43      & U1                                    \\
64    & 0:55       & 5:55 & 14.36 & 28.3                & 1003    & 70.2              & 43      & U1                                    \\
65    & 1:00       & 6:00 & 14.42 & 28.3                & 1003    & 69.3              & 43      & U1                                    \\
66    & 1:05       & 6:05 & 14.58 & 28.5                & 1003    & 68.3              & 43      & U1                                    \\
67    & 1:10       & 6:10 & 14.62 & 28.6                & 1003    & 67.3              & 43      & U1                                    \\
68    & 1:15       & 6:15 & 14.97 & 28.8                & 1003    & 66.3              & 43      & U1                                    \\
69    & 1:20       & 6:20 & 15.33 & 29                  & 1003    & 65.3              & 42      & U1                                    \\
70    & 1:25       & 6:25 & 16.69 & 28.9                & 1003    & 64.2              & 42      & U1                                    \\
71    & 1:30       & 6:30 & 18.33 & 28.8                & 1003    & 63.2              & 42      & U1                                    \\
72    & 1:35       & 6:35 & 18.51 & 28.8                & 1003    & 62.1              & 40      & U1                                    \\
73    & 1:40       & 6:40 & 18.74 & 30.2                & 1003    & 61                & 40      & U1                                    \\
74    & 1:45       & 6:45 & 18.81 & 32.2                & 1003    & 59.9              & 40      & U1                                    \\
75    & 1:50       & 6:50 & 19.56 & 32.2                & 1003    & 58.8              & 40      & U1                                    \\
76    & 1:55       & 6:55 & 19.73 & 30.2                & 1003    & 57.7              & 40      & U1                                    \\
77    & 2:00       & 7:00 & 19.85 & 30.2                & 1003    & 56.6              & 40      & U1                                    \\
78    & 2:05       & 7:05 & 20.95 & 30.3                & 1003    & 55.4              & 40      & U1                                    \\
79    & 2:10       & 7:10 & 21.03 & 30.4                & 1002    & 54.3              & 39      & U2                                    \\
80    & 2:15       & 7:15 & 21.11 & 29.8                & 1002    & 53.2              & 40      & U2                                    \\
81    & 2:20       & 7:20 & 21.13 & 29.8                & 1002    & 52                & 40      & U2                                    \\
82    & 2:25       & 7:25 & 21.11 & 29.7                & 1002    & 50.9              & 40      & U2                                    \\
83    & 2:30       & 7:30 & 21.2  & 29.3                & 1002    & 49.7              & 40      & U2                                    \\
84    & 2:35       & 7:35 & 21.14 & 29.6                & 1002    & 48.6              & 41      & U2                                    \\
85    & 2:40       & 7:40 & 21.15 & 29.8                & 1002    & 47.4              & 40      & U2                                    \\
86    & 2:45       & 7:45 & 21.06 & 29.7                & 1002    & 46.2              & 41      & MAX                                \\
87    & 2:50       & 7:50 & 21.03 & 29.4                & 1002    & 45.1              & 41      & U3                                    \\
88    & 2:55       & 7:55 & 21.19 & 29.2                & 1002    & 43.9              & 42      & U3                                    \\
89    & 3:00       & 8:00 & 21.24 & 29                  & 1002    & 42.7              & 43      & U3                                    \\
90    & 3:05       & 8:05 & 21.22 & 29.1                & 1002    & 41.6              & 45      & U3                                    \\
91    & 3:10       & 8:10 & 21.15 & 28.6                & 1003    & 40.4              & 45      & U3                                    \\
92    & 3:15       & 8:15 & 21.22 & 28.4                & 1003    & 39.2              & 46      & U3                                    \\
93    & 3:20       & 8:20 & 21.22 & 28.3                & 1003    & 38.1              & 46      & U3                                    \\
94    & 3:25       & 8:25 & 21.26 & 28.1                & 1002    & 36.9              & 44      & U3                                    \\
95    & 3:30       & 8:30 & 21.19 & 28.4                & 1002    & 35.7              & 46      & U4                                    \\
96    & 3:35       & 8:35 & 21.12 & 28.7                & 1002    & 34.5              & 47      & U4                                    \\
97    & 3:40       & 8:40 & 20.59 & 28.9                & 1001    & 33.3              & 46      & U4                                    \\
98    & 3:45       & 8:45 & 20.01 & 29.3                & 1001    & 32.2              & 47      & U4                                    \\
99    & 3:50       & 8:50 & 19.39 & 29.2                & 1001    & 31                & 47      & U4                                    \\
100   & 3:55       & 8:55 & 18.9  & 29.2                & 1001    & 29.8              & 47      & U4                                    \\
101   & 4:00       & 9:00 & 18.74 & 29.2                & 1001    & 28.6              & 47      & U4                                    \\
102   & 4:05       & 9:05 & 18.34 & 29.1                & 1001    & 27.4              & 41      & U4                                    \\
103   & 4:10       & 9:10 & 18.04 & 29.1                & 1002    & 26.2              & 42      & U4                                    \\
104   & 4:15       & 9:15 & 17.88 & 28.7                & 1002    & 25.1              & 42      & U4                                    \\
105   & 4:20       & 9:20 & 17.79 & 28.7                & 1002    & 23.9              & 42      & U4                                    \\
106   & 4:25       & 9:25 & 17.71 & 28.5                & 1004    & 22.7              & 42      & U4                                    \\
107   & 4:30       & 9:30 & 17.67 & 28.5                & 1004    & 21.5              & 42      & U4                                    \\
108   & 4:35       & 9:35 & 17.63 & 28.3                & 1004    & 20.3              & 42      & U4                                    \\
      &            &      &       &                     &         &                   &         &                                          \\  \hline                        
\end{tabular}
}
\end{table}

\section{Analysis and Results}
\label{sec:results}

\begin{figure}[!t]
  \includegraphics[width=\columnwidth]{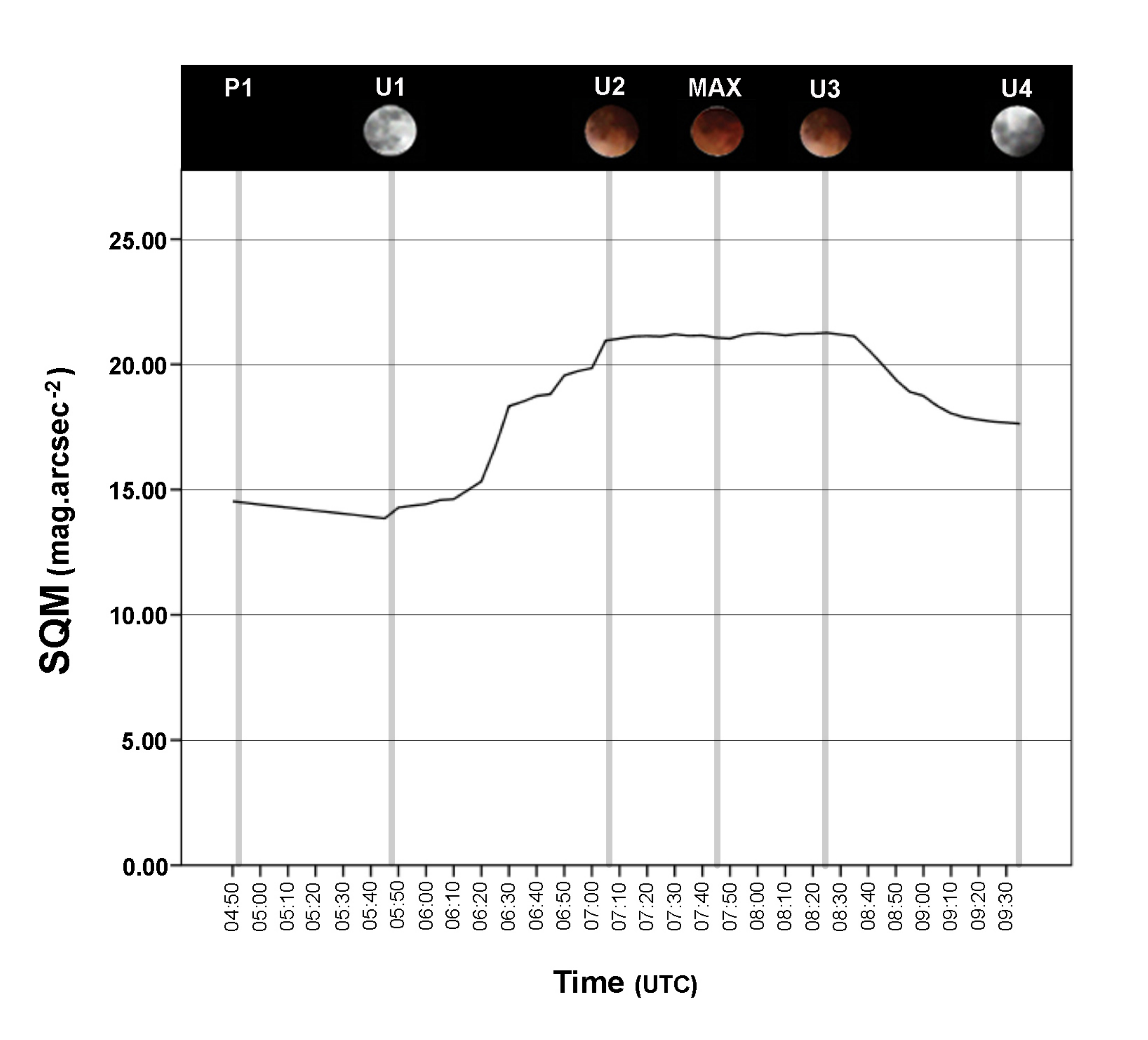}
  \caption{Plot displaying the time evolution of the sky brightness during the total lunar eclipse observed in Colombia on 14-15 April 2014. Upper image shows the different stages as listed in Table~\ref{tab:1}). Note that times are shown in UTC.}
  \label{fig:5}
\end{figure}

Figure~\ref{fig:5} displays the temporal variation of sky brightness as plotted from all acquired data. The multiple total lunar eclipse stages are identified at the corresponding times (see Table~\ref{tab:1}).  The plot shows clear periods of stability, increase and decrease in the sky brightness, as follows: from P1 to U1 there are not many variations, but starting from U1 there is a continuous increment ending at U2. This moment marks the beginning of the totality phase, in which the sky brightness remains steady for all the period, that gets extended until U3 is reached. Soon after, the trend of the plot reveals, for the first time, a diminishing of sky brightness. 

Detailed analysis of figure~\ref{fig:5} allows to establish an average SQM value of 21.15 mag.arcsec$^{-2}$ during the totality phase of the eclipse that can be associated to a Bortle Class 4 of sky quality, as itemized in Table~\ref{tab:3} with the international measurement scale. Up to this moment, and from the beginning of the eclipse, SQM values increased and therefore the sky quality improved \citep{Rabaza}, meaning that during this period, the sky can be classified in a Bortle class ranging from 1 to 8.9.

\begin{table}[!t]\centering
  \setlength{\tabnotewidth}{0.5\columnwidth}
  \tablecols{3}
  \setlength{\tabcolsep}{2.8\tabcolsep}
  \caption{Scale measurements of the sky quality} \label{tab:3}
 \begin{tabular}{cc}
    \toprule
    Sky brightness (mag.arcsec$^{-2})$& \multicolumn{1}{c}{Bortle Class} \\
    \midrule
    $>$21.90 & 1   \\
    21.90 - 21.50 & 2\\
    21.50 - 21.30 & 3\\
    21.30 - 20.80 & 4\\
    20.80 - 20.10 & 4.5\\
    20.10 - 19.10  & 5\\
    19.10 - 18.00 & 6.7 \\
    $<$18.00 & 8.9 \\
    \bottomrule
  \end{tabular}
\end{table}

\begin{figure}[!t]
  \includegraphics[width=\columnwidth]{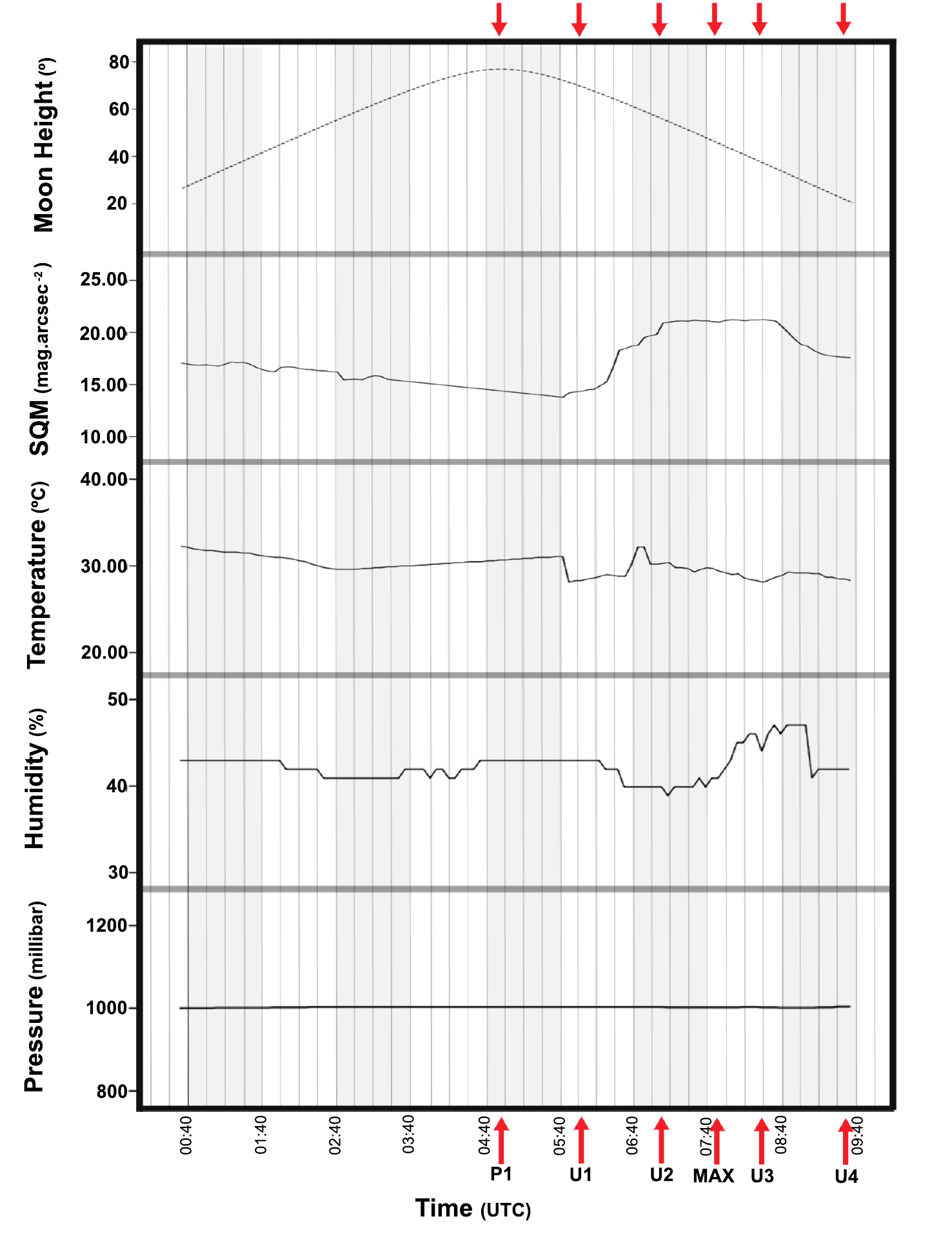}
  \caption{Temporal evolution of physical parameters (Moon height, SQM, temperature, humidity, pressure) during the total lunar eclipse observed in Colombia on 14$^{th}$-15$^{th}$ April 2014. The red arrows on top stand as a reference of the times pointed by the bottom arrows, for stages P1, U1, U2, MAX, U3 and U4, respectively. Note that times are shown in UTC.}
  \label{fig:6}
\end{figure}

\begin{figure}[!t]
  \includegraphics[width=\columnwidth]{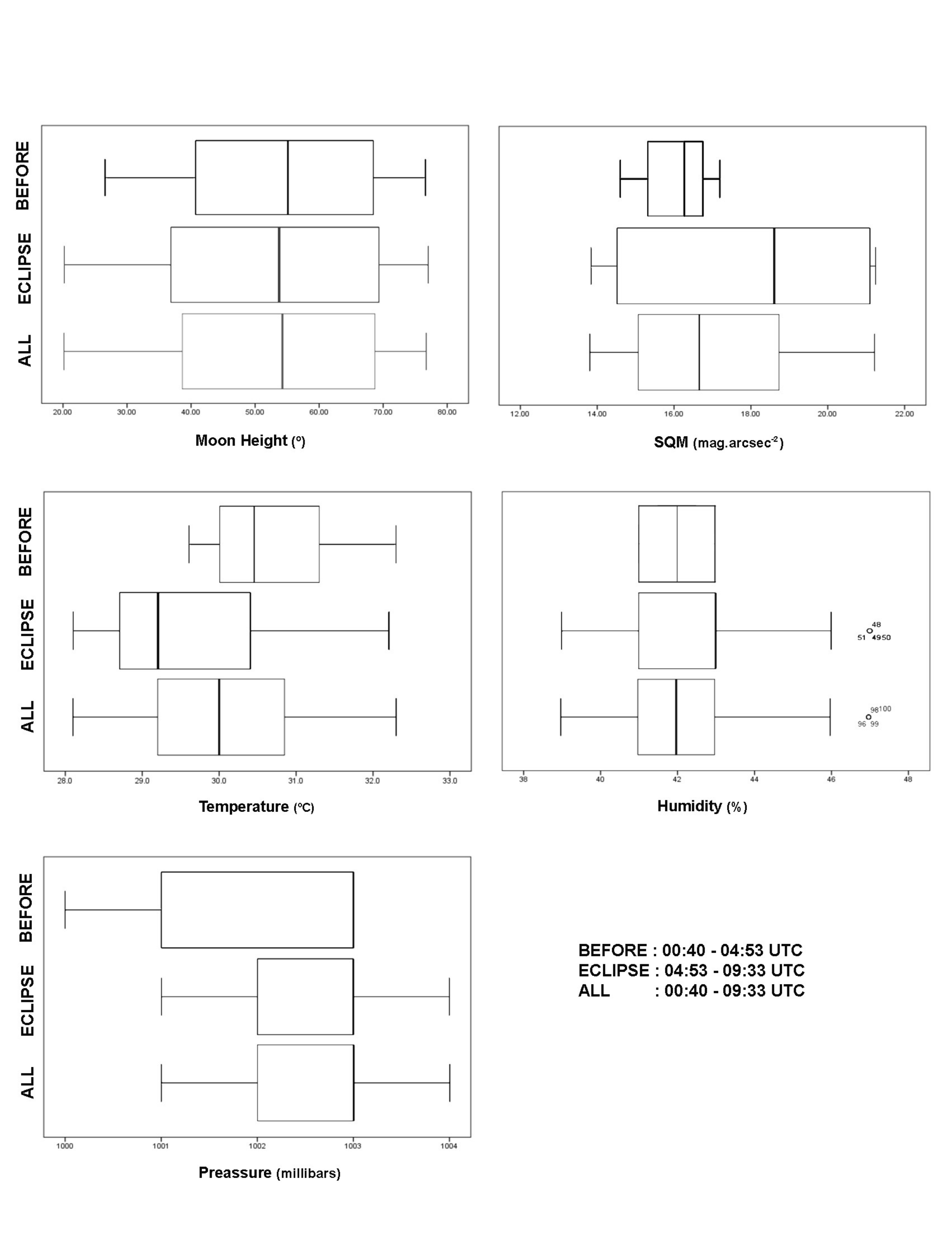}
  \caption{Statistical box plots for several physical parameters {\bf before} the occurrence of the eclipse (BEFORE), during the eclipse (ECLIPSE) and for all the acquired data (ALL) on 14$^{th}$-15$^{th}$ April 2014.  Every plot shows the median value (vertical solid line) , $\pm 1 \sigma$ and total range. Atypical values for humidity are marked by small circles. The mean values and standard deviations are listed in Table~\ref{tab:4}.}   
  \label{fig:7}
\end{figure}

Figure~\ref{fig:6} plots the temporal evolution of different physical parameters. The value of Moon height and sky brightness measurements are also included. In particular we are interested in the variation of pressure, humidity and temperature during the occurrence of the eclipse. As expected, there are not significant or abnormal changes in these variables for about 5 hours (from 00:40 to 05:40 UTC) before phase of the eclipse (U1). There is a slight decrease in the sky quality (SQM) associated with the rising of the Moon in the sky (from $\sim$25$^\circ$ to $\sim$77$^\circ$).  From U1 phase until U2 is reached (07:10 UTC), temperature and humidity exhibit substantial changes, that stay up to phase U3 (08:35 UTC), corresponding to the finishing time for totality. Humidity evidences a peaking value soon after, while maximum temperature of 32.2$^\circ$ is reached closer to U2. On the other hand, minimum temperature occurs in U3, about 40 minutes after the maximum (MAX) was observed. 
After totality, from U3 to U4, changes of these variables are still present, with an overall decrease of  both.\\

Relative humidity experiments a clear decrease after U1, when the eclipse is on its way to totality, but once it is reached, it starts to increase to get to a maximum value soon after U3. The maximum humidity value registered is 47\% and the minimum value is 39$\%$. Regarding pressure values, shown in the bottom panel in figure~\ref{fig:6}, there is a tiny variation of pressure with extreme values of 1002 and 1004 millibars. During the totality phase (U2 to U3) the average value registered is 1002 millibars.  Finally, during the maximum of the total lunar eclipse (07:45:40 UT) the Moon height was 46$^\circ$  and the azimuth was 251$^\circ$ from the observation point.

Statistical analysis was carried out in order to study the behavior of all different measurements before and during the eclipse, as shown in figure~\ref{fig:7} from the values listed in table~\ref{tab:4}. It also includes de statistics for all the acquired data.

\begin{table}[]
\centering
\caption{Descriptive Statistics} \label{tab:4}
\resizebox{\textwidth}{!}{%
\begin{tabular}{lcccccclcccccclcccccclcccccc}
\hline
 & \multicolumn{1}{c}{N} & \multicolumn{1}{c}{MINIMUM} & \multicolumn{1}{c}{MAXIMUM} & \multicolumn{2}{c}{MEAN} & \multicolumn{1}{c}{STANDARD DEVIATION} \\ 
 & \multicolumn{1}{c}{} & \multicolumn{1}{c}{} & \multicolumn{1}{c}{} & \multicolumn{1}{c}{Value} & \multicolumn{1}{c}{Standard error} & \multicolumn{1}{c}{} \\ \hline
 
 \hline
ALL DATA (00:40 - 09:35 UT) \\ \hline
SQM & 58 & 13.85 & 21.26 & 18.0598 & 0.37640 & 2.86654 \\ 
Temperature & 58 & 28.1 & 32.2 & 29.552 & 0.1383 & 1.0532 \\ 
Pressure & 58 & 1001 & 1004 & 1002.53 & 0.099 & 0.754 \\ 
Humidity & 58 & 39 & 47 & 42.64 & 0.284 & 2.166 \\ 
Height & 58 & 20.37 & 76.92 & 52.4621 & 2.36154 & 17.98499 \\ \hline

\hline
 BEFORE THE ECLIPSE   (00:40 - 04:45 UT) \\ \hline
SQM & 50& 14.59 & 17.20 & 16.0244 & 0.11687 & 0.82636 \\ 
Temperature & 50 & 29.5 & 32.3 & 30.618 & 0.1132 & 0.8004 \\ 
Pressure & 50 & 1000 & 1003 & 1002.16 & 0.157 & 1.113 \\ 
Humidity & 50 & 41 & 43 & 42.06 & 0.119 & 0.843 \\ 
Height & 50 & 26.55 & 76.73 & 54.2426 & 2.21850 & 15.68714 \\ \hline

\hline
\hline
 DURING THE ECLIPSE  (04:50 - 09:35 UT) \\ \hline
SQM & 108 & 13.85 & 21.26 & 17.1175 & 0.23033 & 2.39368 \\ 
Temperature & 108 & 28.1 & 32.3 & 30.045 & 0.1041 & 1.0815 \\ 
Pressure & 108 & 1000 & 1004 & 1002.36 & 0.092 & 0.952 \\ 
Humidity & 108 & 39 & 47 & 42.37 & 0.164 & 1.705 \\ 
Height & 108 & 20.37 & 76.92 & 53.2864 & 1.62674 & 16.90561 \\ \hline

\end{tabular}%
}
\end{table}

\section{Conclusions}
\label{sec:conclusions}

The expedition to the Tatacoa Desert, in Colombia, to observe the total lunar eclipse on 14$^{th}$-15$^{th}$ April 2014, was successfully accomplished. The team managed to complete 9 hours of continuous observation and registration of meteorological variables and sky brightness.  The month of April is characterized by the occurrence of high rainfalls in the Colombia. Fortunately the conditions presented during the total lunar eclipse in the selected location allow the expedition team to observe and to record, that was not the case for some other observing locations with failed attempts. Nevertheless, from 09:35 UTC onwards, cloudiness presented at the observation area in the Tatacoa did not allowed to take measurements during the last part of the astronomical event, in the final penumbral phase. These conditions might have influenced the humidity and temperature values acquired close to the ending phase of the partiality, and future work will have to manage to registered this data for the whole duration of the eclipse after P4 was reached.

As explained in Section~\ref{sec:results}, it was possible to evidence changes in these variables during the occurrence of the eclipse, except pressure measurements which were very stable for all recorded data. Particularly for the totality phase, temperature and humidity experiment significant variations as seen in Figure~\ref{fig:6}. The corresponding average of sky brightness during totality gives an SQM value of 21.15  mag.arcsec$^{-2}$ and conditions are such as to classify the sky of the Tatacoa Desert according to the Bortle scale as 4.  The maximum total lunar eclipse took place at 07:45:40 UTC on 15$^{th}$ of April, 2014, the moment where the measured temperature was 29.7$^\circ$. At the initial phase of the lunar eclipse, from P1 to U1, the brightness of the sky slightly increased, therefore reaching a minimum SQM value of 13.85 mag.arcsec$^{-2}$. Once entering the umbral phase, the brightness of the sky begins to decrease considerably as expected from a reduction in the illumination of the lunar disc. It can be evidenced by a clear trend  as shown in Figure~\ref{fig:5}. At the precise moment when the eclipse sequence surpasses to the umbral phase, SQM values start to increase. From U2 to U3 the values reached a {\it plateau} that remains stable for almost 1.5 hours.
It is observed that the lowest value recorded by the SQM during the entire eclipse (13.85  mag.arcsec$^{-2}$) corresponds to 8.9 according to the Bortle scale, whereas the highest value was 4. Sources of error in the sky brightness come from the error introduced by the SQM measurements, which typical value is 0.2302, indicating that the samples were taken as accurately as possible based on the previous calibration of the equipment. Other external factors that can influence the local measurements and that should be beared in mind, are cloudiness, latitude from where it is observed, the environment, vegetation, and the time of year.

Close to the moment of maximum altitude registered by the Moon (77$^\circ$), the humidity and pressure do not vary significantly. Humidity is about 43$\%$ with a pressure of 1003 mbar. Pressure values keep very stable throughout the whole observing session and $\sim$75$\%$ of the data collected for atmospheric pressure have variations of only 1 mbar. The pressure was the variable that registered the least variation compared to other variables. There. was no noticeable incidence on the pressure due to the height of the Moon, and the humidity value at that time neither presented a significant change.


When the Moon enters the umbra, the twilight zone generated by Earth, it does not reflect the same amount of light from the Sun and therefore a decrease in temperature is plausible to occur.  In our records, the ambient temperature varies depending on the transition zone the Moon reaches. In P1 the temperature is higher, regarding averages for the penumbra (U1) and umbra (U2), in which lower temperatures are registered,.Nevertheless, in this period of time we evidenced some clouds approaching to the location, and therefore the temperature decrease can not be directly associated to the totality phase of the eclipse. Further studies should be made to clarify the actual dependance of the radiation from the Sun reflected by the Moon diminished during the totality phase and the ambiente temperature reduction. We can only determine that  maximum humidity coincides with a minimum temperature value.

Regarding the quality of the sky, depending on the levels of sky brightness, it was observed that SQM values start increasing when the eclipse is in phase U1 with a maximum value of 21.26 mag.arcsec$^{-2}$. It means that the quality of the sky improves considerably because there is less brightness of the sky and therefore, the sensor captures fewer  photons. It is also discernible from the Moon height and SQM plots in Figure~\ref{fig:5} that the sky brightness at the beginning (00:40 UTC) and the end (09:35 UTC) of the observing run are very similar with SQM values of 17.11 and 17.63 mag.arcsec$^{-2}$, respectively, with also a rather similar Moon height values of 26.6$^\circ$ and 20.3$^\circ$ for every case. This confirms that the effect of sky quality variation depends almost entirely on the eclipse occurrence and not on the altitude of the Moon over the horizon and the way it affects the scattered light in the celestial vault. Unfortunately, as mentioned before, the weather conditions were not favorable after U4 to continue acquiring data and measure the conditions after P4 was reached. Nevertheless, it can be established without further inspection, that the sky quality of that night at the Tatacoa Desert corresponded to a Bortle Class  value of 8.9 when the eclipse was not in action.

Figure~\ref{fig:7} displays statistical box plots for all the different parameters separated in three different intervals:  before the occurrence of the eclipse (BEFORE), during the eclipse (ECLIPSE) and for all the acquired data (ALL), in order to quantify the influence of the eclipse on the normal conditions of at the beginning of the night, and how median values are affected. 

Our results were are compared with recent studies based on comparative data,  reports, and measurements associated with the extension project of the National Astronomical Observatory of Colombia entitled "Monitoring the quality of the sky in the Tatacoa desert to obtain the Starlight certification as an astrotourism destination\footnote{Visit the information about this project: http://168.176.14.11/index.php?id=6181}  \citep{Pinzon2020}. The study included measurements obtained during the new moons over a year, from January to December 2018,  and their respective analysis, confirming that the quality of the Tatacoa sky is compatible with the requirements as a Starlight astronomical destination\footnote{Website: www.fundacionstarlight.org}  In this comparative study aims at monitoring the quality of the sky in three locations to characterize this destination, the authors recorded their data at the beginning, in the middle, and at the end of the night, in units of  mag.arcsec$^{-2}$ verifying that, according to their results, the required values are close to, or greater than 21 mag.arcsec$^{-2}$, in agreement with the the result for sky quality presented in our work during the totality phase of a lunar eclipse.

\end{document}